\documentclass[fleqn,usenatbib]{mnras}

\usepackage{newtxtext,newtxmath}

\usepackage[T1]{fontenc}
\usepackage{ae,aecompl}

\usepackage{graphicx}	
\usepackage{amsmath}	
\usepackage{amssymb}	
\usepackage{soul}
\soulregister\citet7
\soulregister\citep7
\soulregister\ref7

\title[The curious case of the `fastest' star in \textit{Gaia} DR2]{Lessons from the curious case of the `fastest' star in \textit{Gaia} DR2}

\author[D. Boubert et al.]{
D. Boubert,$^{1}$\thanks{E-mail: douglas.boubert@magd.ox.ac.uk (DB)}
J. Strader,$^{2}$
D. Aguado,$^{3}$
G. Seabroke,$^{4}$
S. E. Koposov,$^{5,3}$
J. L. Sanders,$^{3}$
\newauthor
$\;$S. Swihart,$^{2}$
L. Chomiuk,$^{2}$
and N. W. Evans$^{3}$
\\
$^{1}$Magdalen College, University of Oxford, High Street, Oxford OX1 4AU, UK\\
$^{2}$Department of Physics and Astronomy, Michigan State University, East Lansing, MI 48824, USA\\
$^{3}$Institute of Astronomy, University of Cambridge, Madingley Road, Cambridge CB3 0HA, UK\\
$^{4}$Mullard Space Science Laboratory, University College London, Dorking, Surrey RH5 6NT, UK\\
$^{5}$McWilliams Center for Cosmology, Carnegie Mellon University, 5000 Forbes Ave, Pittsburgh, PA 15213, USA\\
}

\date{Accepted XXX. Received YYY; in original form ZZZ}

\pubyear{2015}

\begin{document}
\label{firstpage}
\pagerange{\pageref{firstpage}--\pageref{lastpage}}
\maketitle

\begin{abstract}
\textit{Gaia} DR2 5932173855446728064 was recently proposed to be unbound from the Milky Way based on the $-614.3\pm2.5\;\mathrm{km}\;\mathrm{s}^{-1}$ median radial velocity given in \textit{Gaia} DR2. We obtained eight epochs of spectroscopic follow-up and find a very different median radial velocity of $-56.5 \pm 5.3\;\mathrm{km}\;\mathrm{s}^{-1}$. If this difference were to be explained by binarity, then the unseen companion would be an intermediate-mass black hole; we therefore argue that the \textit{Gaia} DR2 radial velocity must be in error. We find it likely that the spectra obtained by \textit{Gaia} were dominated by the light from a star $4.3\;\mathrm{arcsec}$ away, and that, due to the slitless, time delay integration nature of \textit{Gaia} spectroscopy, this angular offset corresponded to a spurious $620\;\mathrm{km}\;\mathrm{s}^{-1}$ shift in the calcium triplet of the second star. We argue that such unanticipated alignments between stars may account for 105 of the 202 stars with radial velocities faster than $500\;\mathrm{km}\;\mathrm{s}^{-1}$ in \textit{Gaia} DR2 and propose a quality cut to exclude stars that are susceptible. We propose further cuts to remove stars where the colour photometry is suspect and stars where the radial velocity measurement is based on fewer than four transits, and thus produce an unprecedentedly clean selection of \textit{Gaia} RVS stars for use in studies of Galactic dynamics.
\end{abstract}




\section{Introduction}

The fastest stars travel so rapidly that they can escape from the Milky Way's gravitational well. These `hypervelocity stars' are objects of on-going study because their origin in the tidal disruption of binaries by massive black holes, chaotic N-body stellar encounters or the supernova of their companion star makes them invaluable probes of those events. Only a few tens of hypervelocity stars are known\footnote{See The Open Fast Stars Catalog \citep{2018MNRAS.479.2789B} at \url{https://faststars.space} for an up-to-date listing of hypervelocity stars.} and thus we are still in the regime where individual hypervelocity star discoveries are noteworthy. 

The \textit{Gaia} space telescope \citep{2016A&A...595A...1G} is expected to enable the discovery of hundreds of hypervelocity stars \citep{2018MNRAS.476.4697M}, because it will make the first measurements of the tangential velocities of over a billion stars and the radial velocities of over 100 million. The diverse data produced by \textit{Gaia} have necessitated separate pipelines for the astrometry, Radial Velocity Spectrometer (RVS) and colour photometry ($G_{\mathrm{BP}}$  and $G_{\mathrm{RP}}$), and each of these complicated pipelines were still in active development at the time of the preliminary second \textit{Gaia} data release (DR2, \citealp{2018A&A...616A...1G}) which limited the number of stars published with these measurements. Nonetheless, \citet{2018MNRAS.476.4697M} predicted that among the limited sample of 7,224,631 stars with radial velocities in DR2 there would be a handful of hypervelocity stars.

The \textit{Gaia} DR2 RVS sample was a crowded hunting ground with \citet{2018MNRAS.tmp.2466M}, \citet{2018ApJ...866..121H} and \citet{2018ApJ...868...25B} all conducting searches for high-velocity stars:
\begin{itemize}
\item \citet{2018MNRAS.tmp.2466M} identified as many as 20 stars with a probability greater than 80\% of being unbound from the Milky Way. Surprisingly, only 7 of these stars were consistent with originating in the Milky Way disk and thus \citet{2018MNRAS.tmp.2466M} proposed an extragalactic origin for the remaining 13 stars.
\item \citet{2018ApJ...866..121H} reported 30 stars with extreme space velocities (greater than $480\;\mathrm{km}\;\mathrm{s}^{-1}$). They conjectured that one or two could originate in the Galactic centre, as many as three might originate in the LMC, and that the remaining stars were likely halo objects based on their old age. \citet{2018ApJ...866..121H} noted that this implies the escape velocity near the Sun must be around $600\;\mathrm{km}\;\mathrm{s}^{-1}$.
\item \citet{2018ApJ...868...25B} whittled the \textit{Gaia} DR2 RVS sample down to just 25 likely high-velocity stars and singled out two as being likely hypervelocity stars, while the other high-velocity stars were categorized as statistical outliers. Of their two likely candidates, \citet{2018ApJ...868...25B} cautioned that \textit{Gaia} DR2 1383279090527227264 is possibly a bound late-type giant and that \textit{Gaia} DR2 5932173855446728064 would require follow-up observations because it lies in a crowded field.
\end{itemize}

There are many potential pitfalls when identifying hypervelocity stars in \textit{Gaia} DR2, because calculation of the total Galactocentric velocity relies on estimating a distance from the parallax. There are known issues with the \textit{Gaia} DR2 parallaxes, such as a systematic offset that varies as a function of position and magnitude and the need to add a systematic component to the published uncertainties. If these issues are not accounted for then the distance may be overestimated, which propagates into an inflated Galactocentric velocity and a false positive hypervelocity candidate. One example of this is given in Appendix D of \citet{2018MNRAS.tmp.2466M}, where the authors show that including the approximate global parallax offset of $-0.029\;\mathrm{mas}$ results in only 4 out of 20 of their candidates still having a probability greater than $80\%$ of being unbound. Similarly in Appendix E, \citet{2018MNRAS.tmp.2466M} show that appropriately inflating the uncertainties in the parallax (without including the parallax offset) causes only 5 out of 20 of their candidates to still be likely unbound. Undoubtedly, the uncertainties and systematics will be better understood with the later \textit{Gaia} data releases. Until then, it remains true that the only guaranteed hypervelocity stars are those in which the radial velocity alone is greater than the escape velocity. 

Fortuitously, there is one such star among the crop of candidates found by these three searches: \textit{Gaia} DR2 5932173855446728064 (hereafter referred to as \textit{Gaia} DR2 593...064). This star has an incredible radial velocity of $-614.286\pm2.492\;\mathrm{km}\;\mathrm{s}^{-1}$, which alone is sufficient to class it as a hypervelocity star. This object was the premier candidate of both \citet{2018MNRAS.tmp.2466M} and \citet{2018ApJ...868...25B}. It was absent from the candidate list of \citet{2018ApJ...866..121H} due to their choice to select high-tangential velocity stars as a means to avoid stars with spuriously large radial velocities. Using the methodology of The Open Fast Stars Catalog \citep{2018MNRAS.479.2789B}, we find that the precise parallax of $0.454\pm0.029\;\mathrm{mas}$ places \textit{Gaia} DR2 593...064 at $2.08\pm0.12\;\mathrm{kpc}$. By contrast, \citet{2018MNRAS.tmp.2466M} found $2197_{-120}^{+162}\;\mathrm{pc}$  and \citet{2018ApJ...868...25B} found $2.2\pm0.1\;\mathrm{kpc}$ -- the difference arises because the latter two papers neglected to include the parallax offset. Taking the parallax, proper motions and radial velocity together, the star has a total Galactocentric space velocity of $749.6\pm6.8\;\mathrm{km}\;\mathrm{s}^{-1}$ (\citealp{2018MNRAS.tmp.2466M} found $747_{-3}^{+2}\;\mathrm{km}\;\mathrm{s}^{-1}$, \citealp{2018ApJ...868...25B} found $747\pm3\;\mathrm{km}\;\mathrm{s}^{-1}$; the offset in the medians is likely due to differing choices for the Solar motion and the larger size of our uncertainty is because we propagated the uncertainties in both the location and velocity of the Sun). \citet{2018ApJ...868...25B} commented that the de-reddened colours suggest it is either an A-type main sequence star or in the process of evolving off the main sequence. 

\citet{2018MNRAS.479.2789B} found that the nearest main sequence candidate hypervelocity stars to the Sun are Li10 at $3.2\;\mathrm{kpc}$ \citep{2015RAA....15.1364L} and SDSS J013655.91+242546.0 at $8.5\;\mathrm{kpc}$ \citep{2009JPhCS.172a2009T}, and thus there is a possibility that \textit{Gaia} DR2 593...064 is the nearest known main sequence hypervelocity star to the Sun. The known main-sequence hypervelocity stars have a mean heliocentric distance of more than $50\;\mathrm{kpc}$ \citep{2018MNRAS.479.2789B} and thus a hypervelocity star as close as $2\;\mathrm{kpc}$ would probe a new kinematic regime and allow the first detailed characterisation of a hypervelocity star's motion, atmosphere and chemistry. \citet{2018ApJ...868...25B} sounded a note of caution because \textit{Gaia} DR2 593...064 was the only one of their 25 candidates that was flagged as a \textsc{duplicated\_source} (likely due to it being in a crowded field) and suggested that follow-up observations would be necessary.

The objective of this work is to conduct follow-up of \textit{Gaia} DR2 593...064 to ascertain whether it is a genuine hypervelocity star. In Sec. \ref{sec:observations}, we highlight the unusual concentration of other stars around \textit{Gaia} DR2 593...064 in SkyMapper images and present eight new epochs of ground-based spectroscopic radial velocities which are in tension with the \textit{Gaia} radial velocity. We argue in Sec. \ref{sec:discussion} that the inconsistency can only be reconciled if either \textit{Gaia} DR2 593...064 is in orbit around an intermediate-mass black hole or the \textit{Gaia} measurement is spurious, and conclude the latter to be much more likely. In the remainder of Sec. \ref{sec:discussion} we identify a novel failure mode in the \textit{Gaia} RVS pipeline and discuss the implications of this for the use of \textit{Gaia} radial velocities.

\section{Observations}
\label{sec:observations}
\subsection{Images and photometry}
\begin{figure}
	\includegraphics[width=\columnwidth]{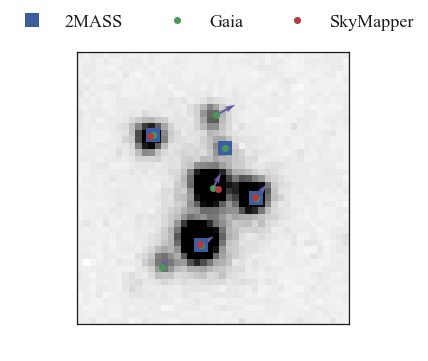}
	\includegraphics[width=\columnwidth]{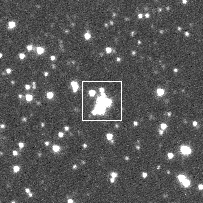}
    \caption{$20\;\mathrm{arcsec}$ and $100\;\mathrm{arcsec}$ cut-outs from SkyMapper images centred on \textit{Gaia} DR2 593...064. Reported photometric sources in 2MASS, \textit{Gaia} and SkyMapper and arrows indicating the magnitude and direction of the \textit{Gaia} proper motion are over-plotted in the top panel.}
    \label{fig:cutouts}
\end{figure}

\textit{Gaia} DR2 593...064 lies close to the plane of the Milky Way at $(l,b)=(329.9^{\circ},-2.7^{\circ})$, which led  \citet{2018ApJ...868...25B} to conjecture that the \textit{Gaia} data may suffer from it being in a crowded field. In the \textit{Gaia} catalogue, there are 18 other sources within a $20\times20\;\mathrm{arcsec}$ cut-out centred on \textit{Gaia} DR2 593...064. To illustrate the density of these sources, we show $20\;\mathrm{arcsec}$ and $100\;\mathrm{arsec}$ images from SkyMapper DR1 \citep{2018PASA...35...10W} in Fig. \ref{fig:cutouts}. We queried the 2MASS point source catalogue \citep{2006AJ....131.1163S}, SkyMapper DR1.1 and the GLIMPSE Source Catalog \citep[contains photometry from the IRAC instrument on Spitzer]{2003PASP..115..953B,2009PASP..121..213C}; 2MASS contains four sources in this field (excluding \textit{Gaia} DR2 593...064), SkyMapper also contains four sources (including \textit{Gaia} DR2 593...064), whilst GLIMPSE contains five sources (including \textit{Gaia} DR2 593...064). The final list of photometry of \textit{Gaia} DR2 593...064 is given in Tab. \ref{tab:photometry}. 

We quantified whether this density of neighbours is unusual by querying in the full \textit{Gaia} DR2 catalogue for neighbours within $8\;\mathrm{arcsec}$ of the 34 \textit{Gaia} DR2 RVS stars with radial velocity greater than $500\;\mathrm{km}\;\mathrm{s}^{-1}$ that meet the quality criteria of \citet{2018MNRAS.tmp.2466M}. \textit{Gaia} DR2 593...064 has the most neighbours with nine. There is one star with 7 neighbours (\textit{Gaia} DR2 5926621184202272256, but none are within $4\;\mathrm{arcsec}$) and the rest have 3 or fewer neighbours. We concluded that \textit{Gaia} DR2 593...064 is in an abnormally crowded field and thus that follow-up spectra are required to verify the \textit{Gaia} radial velocity.

\begin{table}
\centering
\caption{Literature photometry of \textit{Gaia} DR2 593...064\label{tab:photometry}}
\begin{tabular}{cc}
\hline Band  & Value             \\ \hline
$G$     & $13.8104\pm0.0002$ \\
$G_{\mathrm{BP}}$ & $14.2098\pm0.0012$ \\
$G_{\mathrm{RP}}$ & $13.2223\pm0.0012$ \\
$g$     & $13.994\pm0.043$   \\
$r$     & $13.709\pm0.025$ \\
IRAC 3.6     & $12.137\pm0.097$ \\
IRAC 4.5     & $12.014\pm0.093$ \\
IRAC 5.8     & $11.937\pm0.142$ \\
IRAC 8.0     & $11.850\pm0.115$ \\
\hline
\end{tabular}
\end{table}

\subsection{Spectra of \textit{Gaia} DR2 593...064}

We observed \textit{Gaia} DR2 593...064 on eight epochs over the date range $5^{\mathrm{th}}$ May 2018 to $16^{\mathrm{th}}$ September 2018 using the Goodman Spectrograph \citep{2004SPIE.5492..331C} on the SOAR telescope. Each spectrum was obtained with a 0.95\arcsec\ longslit. The first six epochs used a $1200\;\mathrm{lines}\;\mathrm{mm}^{-1}$ grating, giving a wavelength coverage of 4300--5585 \AA$\;$ with a resolution of 1.7 \AA. The latter two epochs used a $2100\;\mathrm{lines}\;\mathrm{mm}^{-1}$ grating that gave a higher resolution of 0.8 \AA$\;$ over a wavelength range of 4500--5160 \AA$\;$ or 6040--6610 \AA, respectively. The exposure times ranged from 300-600 s depending on conditions. All data were reduced and optimally extracted in the usual way, and wavelength calibrated using contemporaneous FeAr arcs. Barycentric radial velocities were derived through cross-correlation with standards taken with the same setup and these were then corrected to obtain the heliocentric velocities (see Table \ref{tab:rvs}).

\begin{table}
	\centering
	\caption{New ground-based heliocentric radial velocities for \textit{Gaia} DR2 593...064 compared to the median value published in \textit{Gaia} DR2\label{tab:rvs}.}
	\begin{tabular}{rr}
		\hline Julian Date & Radial Velocity\\ 
		& $(\mathrm{km}\;\mathrm{s}^{-1})\phantom{ab}$ \\\hline
		2458243.76 & $-70.0 \pm 9.8\phantom{0}$ \\
		2458278.58 & $-43.3 \pm 10.0$ \\
		2458288.50 & $-59.0 \pm 9.6\phantom{0}$  \\
		2458289.46 & $-54.7 \pm 9.0\phantom{0}$  \\
		2458309.58 & $-82.5 \pm 8.2\phantom{0}$  \\
		2458322.45 & $-46.7 \pm 8.2\phantom{0}$  \\
		2458372.51 & $-58.2 \pm 8.5\phantom{0}$  \\
		2458377.50 & $-52.3 \pm 3.9\phantom{0}$ \\ \hline
		Our median & $-56.5\pm5.3\phantom{0}$ \\
		\textit{Gaia} median & $-614.3\pm2.5\phantom{0}$\\ \hline
		Difference & $\phantom{-}557.8\pm5.9\phantom{0}$ \\ \hline
	\end{tabular}
\end{table}

While \textit{Gaia} DR2 reports a remarkably precise radial velocity of $614.286\pm2.492\;\mathrm{km}\;\mathrm{s}^{-1}$, these two numbers are actually the median of the seven individual radial velocity measurements and the uncertainty on that median, estimated as
\begin{equation}
\epsilon_{v_r} = \sqrt{\left(\sqrt{\frac{\pi}{2N}}S(v_r^t)\right)^2+0.11^2}\label{eq:stderror}
\end{equation}
where $S(v_r^t)$ is the standard deviation of the measurements and the $0.11\;\mathrm{km}\;\mathrm{s}^{-1}$ is the calibration floor (\citealp{2018arXiv180409372K}). Inverting this, we can calculate that the standard deviation of the measurements is $5.17\;\mathrm{km}\;\mathrm{s}^{-1}$. If we calculate the median of our eight epochs in an equivalent way (without including a calibration floor and neglecting the individual measurement uncertainties), then we find $-56.5\pm5.3\;\mathrm{km}\;\mathrm{s}^{-1}$. The median of our radial velocity measurements is offset by $557.8\;\mathrm{km}\;\mathrm{s}^{-1}$ from the \textit{Gaia} median and this difference must come from either the system being a binary (and thus the radial velocity measured by \textit{Gaia} is not the systemic radial velocity) or the \textit{Gaia} measurement being spurious. We discuss both scenarios in the following section.

\begin{figure*}
	\includegraphics[width=\linewidth]{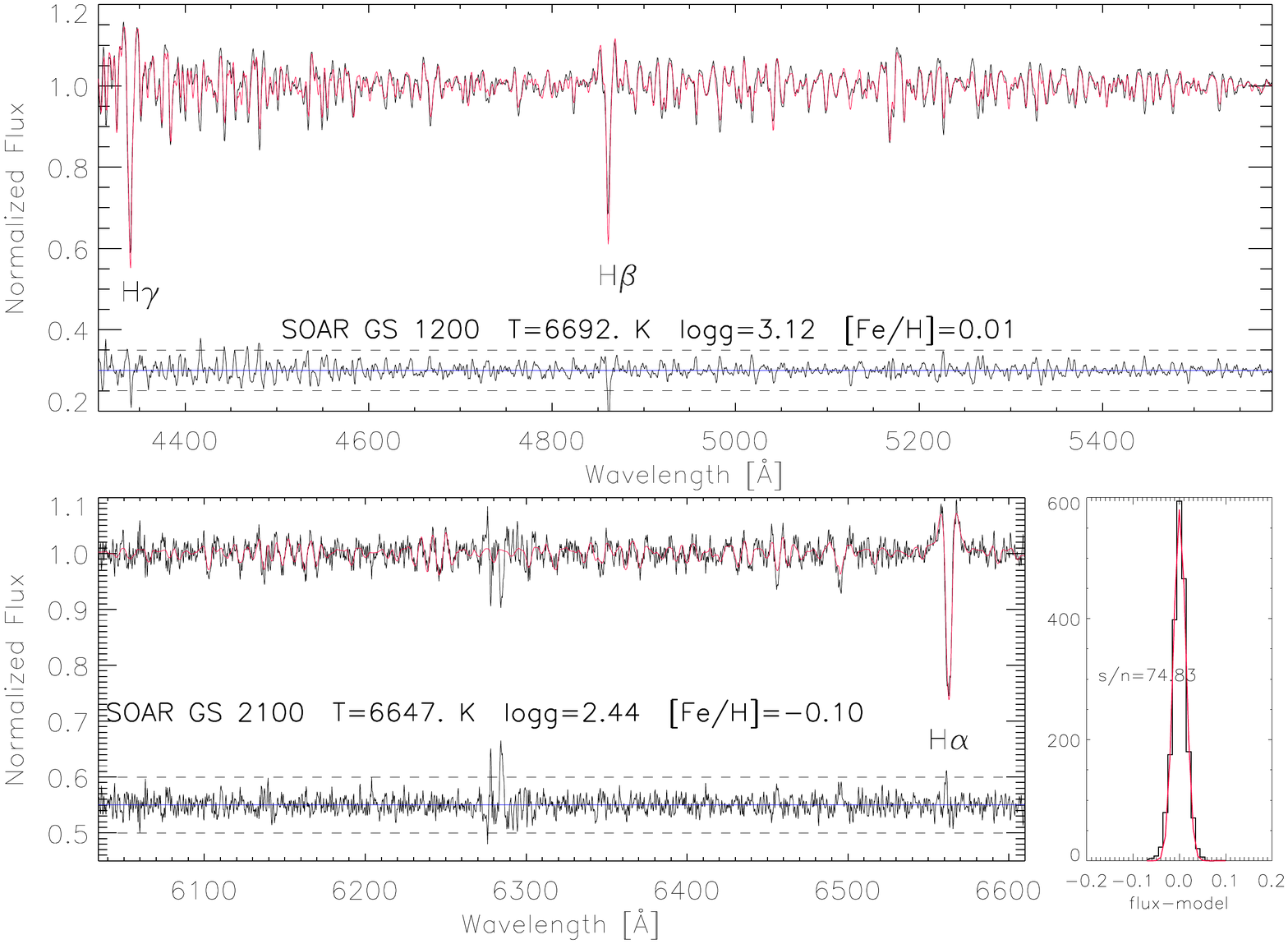}
    \caption{Illustration of the analysis performed on the eight spectra obtained using the Goodman Spectrograph on the SOAR telescope. \textbf{Top:} The normalized stacked spectrum of \textit{Gaia} DR2 593...064 covering the range 4300-5585\,\AA\ (black line) and the best fit derived by FERRE (red line), with the residuals shown at the bottom of the panel. The normalization was done using a running mean filter with a 35 pixel window. The best fit stellar parameters are also shown. \textbf{Bottom left:} Same analysis applied to the normalized spectrum covering the region 6040-6610\,\AA. Note that the series of lines in the region 6270-6300\,\AA\,that are not well fit by the model are likely related to the well-known diffuse interstellar band at around 6284\,\AA. \textbf{Bottom right:} We estimated the signal-to-noise ratio of the spectrum in the top panel by fitting a Gaussian to the residuals (in units of normalized flux), finding $S/N=74.8$.}
    \label{fig:soarplot}
\end{figure*}

To extract information on the atmospheric parameters of \textit{Gaia} DR2 593...064, we use a grid of synthetic spectra computed by \citet{alle18} using the FERRE code \citep{alle06}. The grid contains models with the stellar parameters effective temperature $T_{\rm eff}$, surface gravity $\log g$, and overall metallicity [Fe/H] as free parameters. The limits of the grid are: $-5\leq[\mathrm{Fe}/\mathrm{H}]\leq1$ dex; $5500\leq T_{\mathrm{eff}}\leq8000\;\mathrm{K}$ and $1.0\leq \log g \leq5.0$ dex. In addition, the grid is extended to account for broadening due to rotation ($v\sin i$) in steps of 5\;$\mathrm{km}\;\mathrm{s}^{-1}$. The microturbulence is fixed to $\log{\xi}=0.176\;\mathrm{cm}\;\mathrm{s}^{-1}$. The atmospheric parameters are determined from the Alpha (spectrum covering 6040--6610 \AA), Beta (spectrum covering 4500--5160 \AA) and Blue (a stack of the six spectra with lower resolution) spectra. Both the grid and the three spectra are normalized using a running mean filter with a window of 35 pixels. The FERRE code utilises MCMC to identify a posterior on the atmospheric parameters, assuming cubic interpolation of the spectra between the grid points. Ten Markov chains of 1000 burn-in steps and 4000 science steps were used in each of the three analyses. In Fig. \ref{fig:soarplot}, we illustrate the analysis of the Blue and Alpha SOAR spectra, which have two different gratings ($1200$ and $2100\;\mathrm{lines}\;\mathrm{mm}^{-1}$). 

To aid the discussion in the remainder of this work, we combine the information from the three spectra into a joint posterior on the atmospheric parameters and stellar properties. The method described above results in three sets of MCMC samples from the posterior over the atmospheric parameters, corresponding to the Alpha, Beta and Blue spectra. The Alpha and Blue spectra cover mutually exclusive wavelength regions and thus the inferences drawn from each spectra can be treated as independent, with the implication that the results can be combined to obtain a more precise inference of the atmospheric parameters. Note that the wavelengths covered by the Beta spectrum are a subset of the wavelengths covered by the Blue spectrum and thus we use only the Alpha and Blue spectra to obtain the final atmospheric parameters. 

\begin{figure}
	\includegraphics[width=\columnwidth,clip, trim=0.1cm 0.4cm 0.8cm 1.0cm]{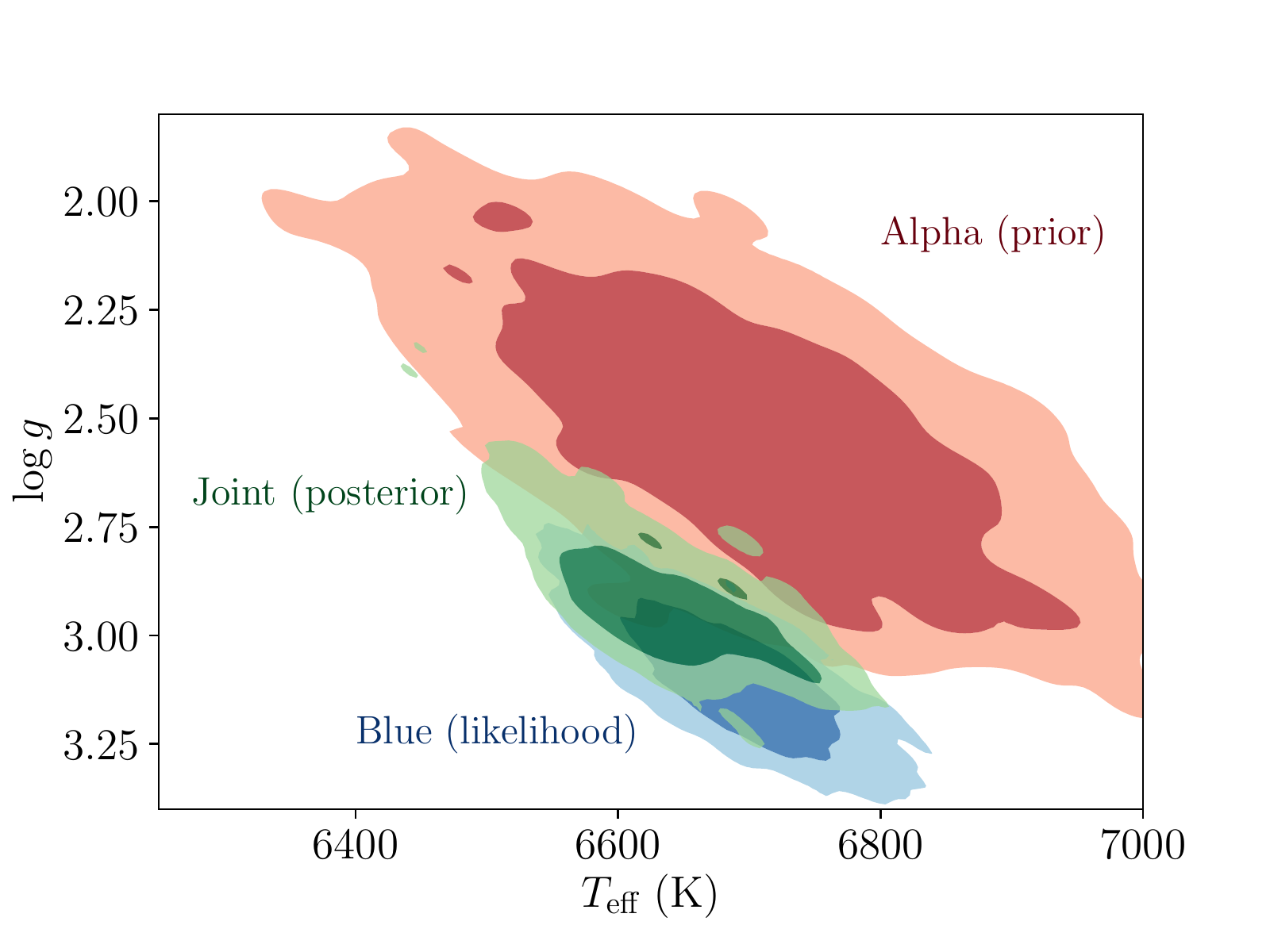}
    \caption{68\% and 95\% contours of the joint posteriors over $\log{g}$ and $T_{\mathrm{eff}}$ based on the Alpha and Blue spectra. When these distributions are treated as the prior and likelihood of a Bayesian inference, the posterior  shown in green is the optimal combined inference. Note that these three distributions are marginalised over $[\mathrm{Fe}/\mathrm{H}]$ for illustrative purposes.}
    \label{fig:post-ast}
\end{figure}

The standard way to combine the inferences drawn from independent datasets is to treat the posterior from applying the model to the first dataset as the prior when applying the model to the second dataset. We note that the two sets of posterior samples can be equivalently viewed as samples from the likelihood, because we assumed a uniform prior on the atmospheric parameters. We use kernel density estimation to construct approximate PDFs describing the two sets of samples and run a third Bayesian inference with the posterior for the Alpha spectrum as the prior and the posterior for the Blue spectrum as the likelihood. One assumption underlying this method is that the kernel density estimate of the PDFs is an accurate representation of the true PDFs, which is only likely to be true if the true PDFs are smooth and continuous (e.g. there is not a valley of low probability lying between the posteriors); the samples describing the Alpha posterior enclose the volume of the Blue posterior, suggesting that this is not the case. Samples are drawn from the posterior using the MCMC \textsc{emcee} \citep{emcee} \textsc{Python} module using 50 walkers with 1000 burn-in steps and 4000 science steps. In Figure \ref{fig:post-ast}, we show this posterior together with the prior and likelihood.

We translate the posterior on $[\mathrm{Fe}/\mathrm{H}]$, $T_{\mathrm{eff}}$ and $\log{g}$ into a posterior on the mass, age, stage, luminosity and radius of the star using the \textsc{PARSEC} version 1.2S isochrones \citep{2017ApJ...835...77M} which features a log-normal IMF and covers the range $6<\log(\mathrm{Age}/1\;\mathrm{yr})<10.1$ with steps of 0.05 and $-2.1<[\mathrm{Fe}/\mathrm{H}]<0.5$ dex with steps of 0.1 dex. We construct a KD-tree of all the isochrone points as a function of $[\mathrm{Fe}/\mathrm{H}]$, $T_{\mathrm{eff}}$ and $\log{g}$ using the \citet{1999cs........1013M} algorithm as implemented in \textsc{SciPy}. For each of the posterior samples obtained in the previous paragraph, we identify the nearest ten isochrone points; the agglomeration of these points can be interpreted as an approximate posterior in the stellar parameters. The medians and $1\sigma$ intervals of the posterior are given in Tab. \ref{tab:spectraisochrones}, while a corner plot of $T_{\mathrm{eff}}$, $\log g$, $[M/H]$ and $v_{\mathrm{rot}}$ is given in Appendix \ref{sec:atmoscorner}. The star is hot ($T_{\mathrm{eff}}=6627_{-46}^{+64}\;\mathrm{K}$) and puffy ($\log g = 2.94_{-0.12}^{+0.08}$) and thus is likely to be an A star that is either on the main sequence or has recently turned off onto the sub-giant branch. The star is rotating at $v\sin{i}=110\pm5\;\mathrm{km}\;\mathrm{s}^{-1}$, which is only slightly faster than the $100\;\mathrm{km}\;\mathrm{s}^{-1}$ median rotation speed in the \citet{2002A&A...393..897R} catalogue of 2151 A-type stars. This rotation rate agrees with the visually broadened lines in the higher resolution spectra.

\begin{table}
	\centering
	\caption{Posterior parameters for \textit{Gaia} DR2 593...064\label{tab:spectraisochrones}}
	\begin{tabular}{cc}
		\hline Parameter & Posterior range            \\ \hline
		$T_{\mathrm{eff}}\;(\mathrm{K})$     & $6627.5_{-46.3}^{+63.6}$ \\[3pt]
		$\log{g}$ & $2.94_{-0.12}^{+0.08}$ \\[3pt]
		$[\mathrm{Fe}/\mathrm{H}]$ & $-0.05_{-0.04}^{+0.03}$ \\[3pt]
		$v\sin i$ & $110_{-5}^{+5}$ \\[3pt]
		Mass $(\mathrm{M}_{\odot})$     & $2.84_{-0.18}^{+0.26}$   \\[3pt]
		Radius $(R_{\odot})$    & $9.42_{-1.26}^{+1.98}$ \\[3pt]
		Age $(\mathrm{Myr})$ & $437_{-98}^{+76}$ \\[3pt]
		Luminosity    & $2.19_{-0.10}^{+0.16}$ \\
		\hline
	\end{tabular}
\end{table}

\subsection{Spectra of nearby stars}
Motivated by the proximity of a number of comparably bright stars (see Fig. \ref{fig:cutouts}), we obtained single-epoch ground-based spectroscopic follow-up of two near neighbours of \textit{Gaia} DR2 593...064: \textit{Gaia} DR2 5932173855446724352 (hereafter \textit{Gaia} DR2 593...352) and \textit{Gaia} DR2 5932173851032088576 (hereafter \textit{Gaia} DR2 593...576). Radial velocities were derived from these spectra using cross-correlation following the same procedure described in the previous section.

\textit{Gaia} DR2 593...352 is not only the brighter of the two nearby stars ($G=13.38$), but it is also brighter than \textit{Gaia} DR2 593...064 ($G=13.81$) and thus is bright enough that it has a \textit{Gaia} DR2 RVS measurement. Our radial velocity of $0.96\pm0.56\;\mathrm{km}\;\mathrm{s}^{-1}$ is entirely consistent with the median RVS value of $5.40\pm2.85\;\mathrm{km}\;\mathrm{s}^{-1}$ based on seven transits. \textit{Gaia} DR2 593...352 is only $4.284\;\mathrm{arcsec}$ away from \textit{Gaia} DR2 593...064 and thus it is curious that \textit{Gaia} DR2 593...352 has neither a $G_{\mathrm{BP}}$ or $G_{\mathrm{RP}}$ reported measurement, which could be an indication that the BP-RP spectra of this star were flagged as blended \citep{2018A&A...616A...3R}.

\textit{Gaia} DR2 593...576 is a $G=14.4$ star only $3.234\;\mathrm{arcsec}$ away from \textit{Gaia} DR2 593...064. It is too faint ($G_{\mathrm{RP}}=13.5$) for it to be surprising that it does not have an RVS measurement in \textit{Gaia} DR2. We find a radial velocity of $-58.62\pm1.41\;\mathrm{km}\;\mathrm{s}^{-1}$, which is consistent with our median radial velocity of \textit{Gaia} DR2 593...064.

We discuss both of these stars in more detail in the following sections.

\section{Discussion}
\label{sec:discussion}

The highly significant discrepancy between the \textit{Gaia} DR2 radial velocity and our own ground-based measurements demands an explanation. We discuss in turn the two scenarios: 1) the \textit{Gaia} measurement is genuine and the system is in a high-amplitude binary or 2) the \textit{Gaia} measurement is spurious in a way that escaped the cuts applied by \textit{Gaia} DPAC. We then discuss the implications of our finding for the handling of \textit{Gaia} DR2 RVS data in future.

\subsection{Scenario 1: the \textit{Gaia} measurement is genuine}
\label{sec:scenario1}

If the \textit{Gaia} measurement is genuine, then the most likely explanation is that \textit{Gaia} DR2 593...064 is in a high-amplitude binary. In this scenario, the radial velocity measured at any given epoch is the systemic velocity of the binary added to the orbital velocity of the star within the binary, and so would vary over the orbital period of the binary. Extreme binaries can have orbital velocities of hundreds of $\mathrm{km}\;\mathrm{s}^{-1}$ (e.g. \citealp{1999AJ....118..515L} and the other 14 papers in that series) and thus this could potentially solve the discrepancy.

\begin{figure}
	\includegraphics[width=\columnwidth,clip, trim=0.3cm 0.4cm 0.4cm 0.4cm]{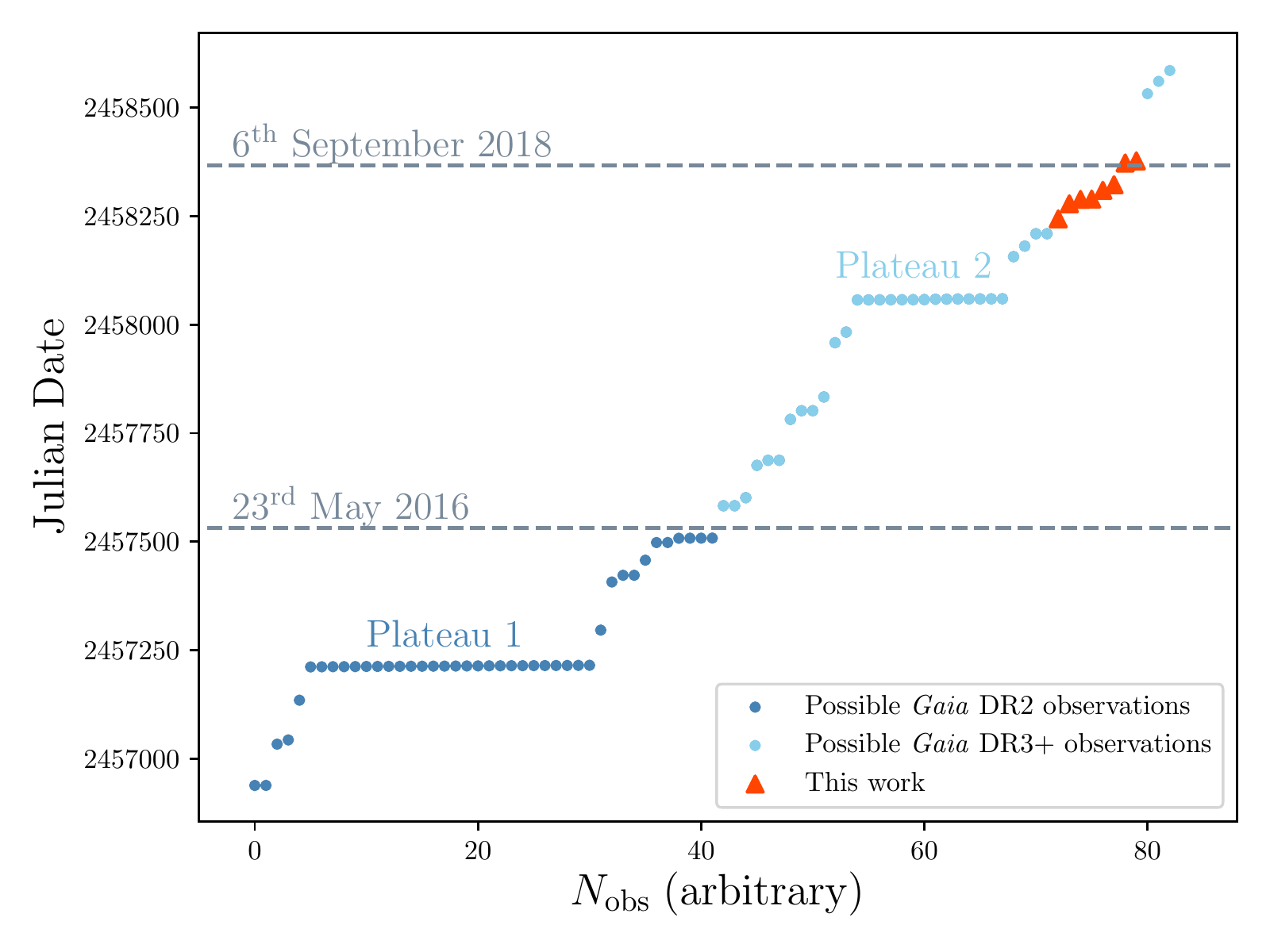}
    \caption{The possible dates on which \textit{Gaia} is predicted to measure the radial velocity of \textit{Gaia} DR2 593...064 and the dates on which our eight epochs were taken. Only data taken prior to 23$^{\mathrm{rd}}$ May 2016 was included in \textit{Gaia} DR2 and of those 42 possible epochs, 26 lie in a `plateau' where it is possible that all seven \textit{Gaia} epochs were obtained in just a few days.}
    \label{fig:rvdates}
\end{figure}

A barrier to testing this hypothesis is that the Julian date of the individual \textit{Gaia} radial velocity measurements was not released in DR2. We obtain a list of predicted dates at which a star at a given location on the sky will transit across CCD rows 4-7 on the \textit{Gaia} focal plane (potentially allowing a radial measurement to be made) from the online ESA Observation Forecast Tool\footnote{\url{https://gaia.esac.esa.int/gost/} accessed 06/09/2018.}. In the full mission \textit{Gaia} DR2 593...064 is predicted to transit CCD rows 4-7 75 times with 42 of these occurring in the time period covered by \textit{Gaia} DR2 (see Fig. \ref{fig:rvdates}). The reason that only seven of these 42 transits resulted in a radial velocity measurement is due to scanning law dead time as a result of `nominal orbital maintenance operations; inadequate resources for placement of the windows at the detection chain level in high-density regions; deletion in the on-board memory as a result of inadequate capacity, particularly when both telescopes are scanning the Galactic plane; and data transmission losses' \citep{2018arXiv180409369C}. The dead time fraction is expected to exceed 40\% at the faint end.

If this is a star in a binary where the measured orbital velocity varies from $-50$ to $-600\;\mathrm{km}\;\mathrm{s}^{-1}$ over more than two years, then it is unlikely that measurements spread over the two-year data gathering period of \textit{Gaia} DR2 would have a standard deviation of only $5\;\mathrm{km}\;\mathrm{s}^{-1}$. Note that 26 of these 42 transits occur over a period of just $3.75\;\mathrm{days}$ (see the region marked as Plateau 1 in Fig. \ref{fig:rvdates}) and if all seven measurements were made in this window and the binary is long-period, it is possible that these seven measurements could be consistent with each other.

How likely is it that the seven transits occurred in this four-day window? If the chance that a transit over CCDs 4-7 results in a successful radial velocity measurement is independent of the orientation of \textit{Gaia}, then the probability of success on any given crossing is independent of the date and thus each transit is independent of the other transits. We can thus say that the probability of all seven successful radial velocities occurring during the plateau is ${26 \choose 7}/{42 \choose 7}=2.4\%$. On the other hand, there is a $93.9\%$ ($45.4\%$) chance that at least three (five) of the transits were during the plateau. However, it is likely that the orientation will matter (as mentioned above, dead time is dependent on whether both telescopes are scanning the Galactic plane) and thus that the probability of a radial velocity being measured is correlated with the date of the transit. Observations made within a 3.75-day window will occur with very similar orientations and thus it is plausible that all seven observations could have occurred during the plateau.

\begin{figure*}
	\includegraphics[width=\linewidth,clip, trim=1.0cm 1.4cm 2.0cm 2.0cm]{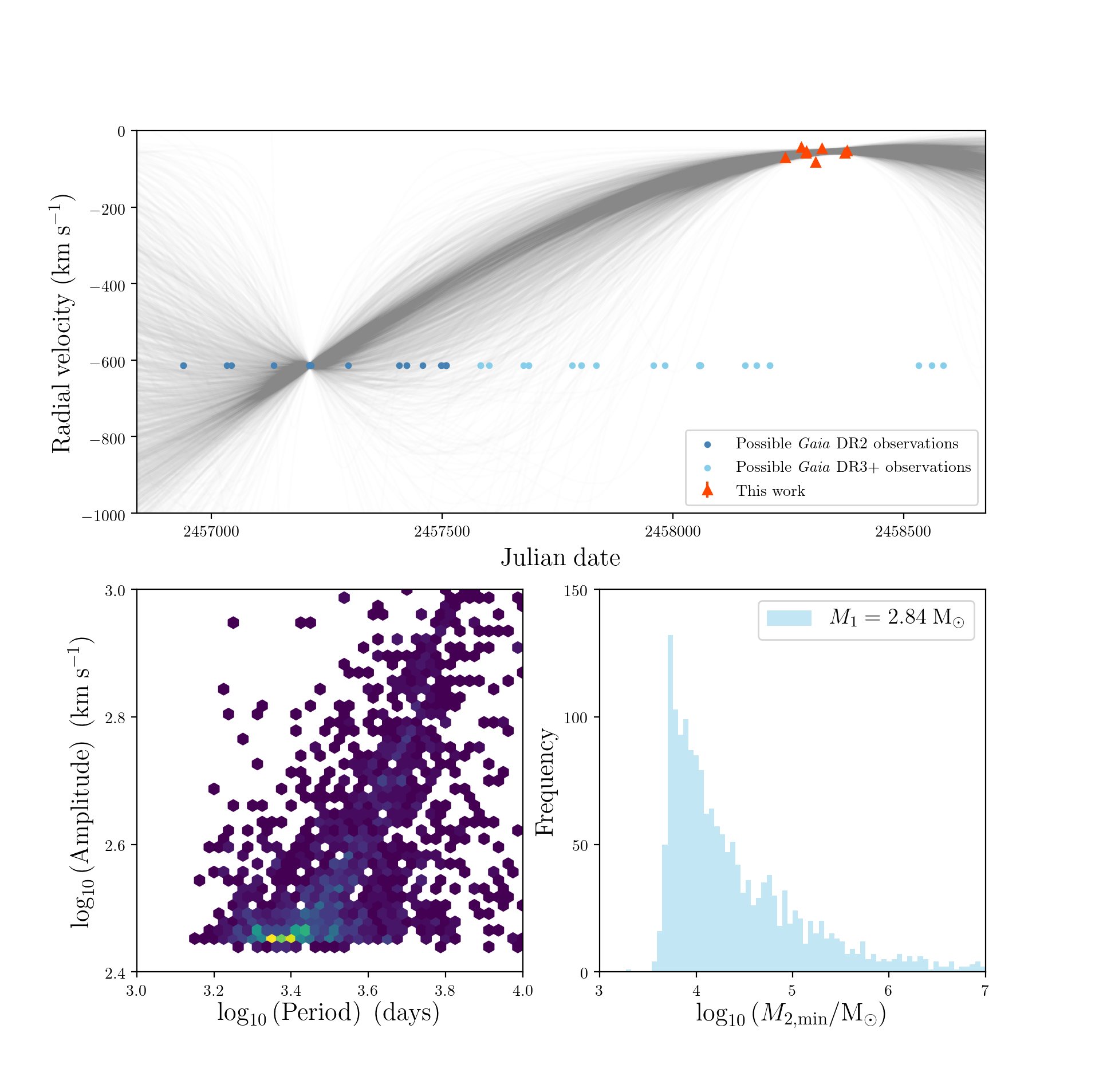}
    \caption{The binary orbital solution if all seven \textit{Gaia} RVS measurements were taken during a 3.75-day period. \textbf{Top:} Posterior radial velocity tracks with the measurements overplotted. \textbf{Bottom left:} Posterior in period-amplitude space. \textbf{Bottom right:} Posterior constraint on the minimum mass of the unseen companion, assuming an inclination of $90\;\mathrm{deg}$.}
    \label{fig:binaryorbitwithgaia}
\end{figure*}

Based on the previous considerations, we assume that all seven \textit{Gaia} RVS transits occurred during the plateau. We randomly choose seven epochs from the 26 epochs during the plateau and assume \textit{Gaia} measured the radial velocity to be $614.29\pm5.17\;\mathrm{km}\;\mathrm{s}^{-1}$ at each of these epochs. Combining these seven epochs with our eight epochs thus gives us a radial velocity time series of 15 points. The question then arises: what are the properties of the fiducial binary system given this dataset? We apply \textsc{The Joker} \citep{Price-Whelan:2017a,2017ApJ...837...20P}, a custom Monte Carlo sampler for the two-body problem, and specify a prior on the orbital period of the binary that is log-uniform over the range $(1,10000)\;\mathrm{days}$. We otherwise use the default priors in \textsc{The Joker}: that the eccentricity follows $\operatorname{Beta}(0.867,3.03)$ \citep{2013MNRAS.434L..51K}, that the pericenter phase and argument are uniform over $[0,2\pi)$, and that the systemic velocity and velocity semi-amplitude are broad Gaussians which are essentially flat over the region of interest. We requested $3\times10^5$ prior samples and obtained $1718$ posterior samples, which we show in Fig. \ref{fig:binaryorbitwithgaia}. The posterior requires that the binary orbital period must be longer than $1000\;\mathrm{days}$ and that the minimum mass of the unseen companion is at least $3\times10^3\;\mathrm{M}_{\odot}$ (calculated by assuming that the binary is edge-on).

\textit{Gaia} DR2 593...064 being a member of a binary system is only a tenable explanation of the large offset between our radial velocities and \textit{Gaia}'s if the unseen companion is an intermediate-mass black hole (IMBH). One circumstantial piece of evidence in favour of this hypothesis is that \textit{Gaia} DR2 593...064 has an unusually large number of neighbours, as shown in Fig. \ref{fig:cutouts} and discussed in Sec. \ref{sec:observations}. \citet{2009ApJ...699.1690M} first proposed that massive black holes $(>10^6\;\mathrm{M}_{\odot})$ ejected from a galaxy centre by the gravitational wave recoil post-merger could host a small cluster of bound stars, which they termed a ``hypercompact stellar system''. \citet{2009MNRAS.395..781O} considered the analogous ejection of the central black holes $(<10^5\;\mathrm{M}_{\odot})$ of the low-mass galaxies that merged to form the Milky Way and found that there may be hundreds of such systems in the halo, typically consisting of tens to hundreds of stars. \citet{2014ApJ...780..187R} used cosmological simulations to predict that hundreds of naked IMBHs (i.e. without an accompanying dark matter subhalo) may populate the Milky Way halo and that the IMBHs within $8\;\mathrm{kpc}$ would host a cusp of stars with an angular extent of $2{-}10\;\mathrm{arcsec}$, comparable to the angular size of the group of stars shown in Fig. \ref{fig:cutouts}.

One argument against an IMBH interpretation is that \textit{Gaia} DR2 593...064 is young ($437_{-98}^{+76}\;\mathrm{Myr}$), while the IMBHs predicted by \citet{2009MNRAS.395..781O} would have been brought into the Milky Way by the minor mergers of small dwarf galaxies and thus should host old stellar populations. Neither of the other two stars of which we obtained spectra exhibit extreme radial velocities (\textit{Gaia} DR2 593..352 at $1.0\pm0.6\;\mathrm{km}\;\mathrm{s}^{-1}$; \textit{Gaia} DR2 593..576 at $-58.6\pm1.4\;\mathrm{km}\;\mathrm{s}^{-1}$) although if they are further out from the nominal black hole then we might not expect them to do so. A strong argument against the existence of the black hole is the close alignment between the radial velocity of \textit{Gaia} DR2 593..576 and the median radial velocity of \textit{Gaia} DR2 593...064 ($-56.5\pm5.3\;\mathrm{km}\;\mathrm{s}^{-1}$); if \textit{Gaia} DR2 593...064 is in orbit around a black hole then this velocity is not at all representative of the systemic velocity of the system, but if it is single (or in a stellar mass binary) then the alignment suggests that these two stars are co-moving. The parallax of \textit{Gaia} DR2 593..576 suggests it is too distant ($\varpi=0.2053\pm0.0364\;\mathrm{mas}$) to be associated with \textit{Gaia} DR2 593...064 ($\varpi=0.4540\pm0.0290\;\mathrm{mas}$), however the significance of its astrometric excess noise is 2.25, indicating that the astrometric solution may not be trustworthy.

One sanity check we performed was to calculate the preferred binary orbital solution for our 8 ground-based measurements on their own. We requested $3\times10^6$ prior samples and obtained $15588$ posterior samples, using the same log-uniform $(1,10000)\;\mathrm{day}$ prior on the period. The motivation for requesting ten times more samples was that many of the posterior samples would require \textit{Gaia} DR2 593...064 to be outside of its Roche lobe. Note that whether a sample $S$ leads to Roche lobe overflow depends on the uncertain mass and radius of \textit{Gaia} DR2 593...217 and on the unknown inclination of the binary. We accounted for this in a probabilistic way:

\begin{enumerate}
    \item We drew 100 random values for the mass and radius of \textit{Gaia} DR2 593...064 from the posterior samples described in Sec. \ref{sec:observations} and, correspondingly, 100 random values for the inclination of the binary.
    \item We determined, for each sampled orbit $S$, the fraction $F$ of the 100 realisations where the Roche lobe radius at periastron \citep{1983ApJ...268..368E} was exterior to the radius of the star.
    \item We drew a uniform variate $u$ for each of the sampled orbits $S$ and discarded that sample if $u$ was greater than $F$.
\end{enumerate}

\begin{figure}
	\includegraphics[width=\columnwidth,clip, trim=0.4cm 1.7cm 4.0cm 5.0cm]{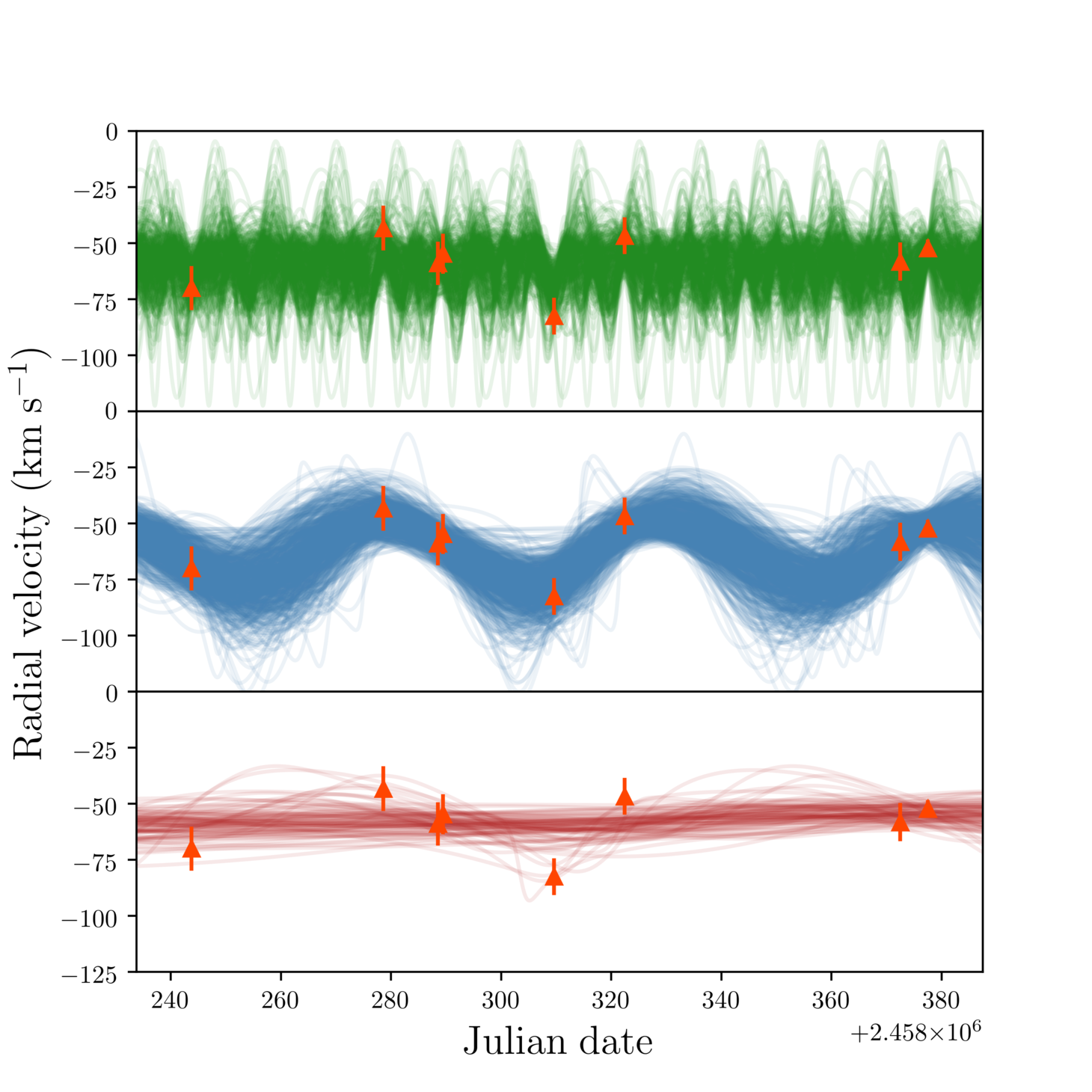}
    \caption{The binary orbital solution based on only our eight ground-based radial velocities. There are three classes of orbits and these are shown in individual panels. \textbf{Top:} Posterior radial velocity tracks for orbits with $P<40\;\mathrm{days}$. \textbf{Middle:} Posterior radial velocity tracks for orbits with $40<P<70\;\mathrm{days}$. \textbf{Bottom:} Posterior radial velocity tracks for orbits with $P>70\;\mathrm{days}$.}
    \label{fig:binaryorbitwithoutgaia}
\end{figure}

This procedure cut the number of orbit samples from $15588$ to $1068$, and thus demonstrated the importance of accounting for the Roche lobe of stars in high-velocity amplitude binary systems. We show the orbits corresponding to these $1068$ samples in Fig. \ref{fig:binaryorbitwithoutgaia}.

There are three classes of solution apparent in Fig. \ref{fig:binaryorbitwithoutgaia}: 30\% have short periods $P<40\;\mathrm{days}$ (top panel of Fig. \ref{fig:binaryorbitwithoutgaia}), 62\% of the orbits lie in a single posterior mode with period $P=51.8\pm4.5\;\mathrm{days}$ (middle panel), and the remainder have long periods $P>70\;\mathrm{days}$ (bottom panel). One interpretation of the long-period orbits is that \textit{Gaia} DR2 593...064 is not in a binary and thus has a constant radial velocity. We tested whether the trend of positive slopes in the bottom panel of Fig. \ref{fig:binaryorbitwithoutgaia} is real by fitting a straight line to the eight radial velocity epochs using \textsc{emcee}; the slope of the line is greater than zero for only $92\%$ of the samples and thus we cannot reject the possibility that \textit{Gaia} DR2 593...064 is single at the $5\%$ level. An argument against this star being single is that there is a strong signal of rotational broadening in the spectral lines ($v\sin i=110\pm5\;\mathrm{km}\;\mathrm{s}^{-1}$) which could be explained by the effect of tides in a close binary. An alternative interpretation of the long-period orbits is that they correspond to the IMBH solution identified above, although this would require that the rapid rotation of \textit{Gaia} DR2 593...064 is explained by a stellar evolutionary process. The short baseline of our eight radial velocity measurements implies that, if the star is in a long-period binary, we cannot accurately constrain the period or velocity amplitude of that binary. In summary, our eight spectroscopic measurements are marginally consistent with the star being single or orbiting an IMBH, but are most consistent with the star being in a close stellar binary. The medium-period solutions imply a secondary mass $M_2$ and separation $a$ of $M_2=1.1_{-0.5}^{+2.3}\;\mathrm{M}_{\odot}$ and $a=94_{-8}^{+16}\;\mathrm{R}_{\odot}$, while the short-period solutions have $M_2=0.4_{-0.2}^{+0.7}\;\mathrm{M}_{\odot}$ and $a=33_{-9}^{+20}\;\mathrm{R}_{\odot}$. The systemic velocity of the medium-period solutions is $-62.5\pm5.0\;\mathrm{km}\;\mathrm{s}^{-1}$.

\subsection{Scenario 2: the \textit{Gaia} measurement is spurious}
\label{sec:scenario2}

The counter scenario is that the radial velocity reported in \textit{Gaia} DR2 is spurious. One could argue that this is unlikely, because quality cuts were applied by \citet{2018A&A...616A...6S} prior to the data release. \citet{2018arXiv180409372K} summarise these cuts as removing:
\begin{itemize}
    \item Stars with an uncertainty on their position greater than $100\;\mathrm{mas}$.
    \item Stars where the $G_{\mathrm{RVS}}$ derived from their RVS spectrum was greater than $14\;\mathrm{mag}$.
    \item Stars where all of their transits are flagged as ambiguous.
    \item Stars with a radial velocity uncertainty greater than $20\;\mathrm{km}\;\mathrm{s}^{-1}$.
    \item Stars where more than 10\% of the transits were detected as being doubly-lined spectroscopic binaries.
    \item Stars which were detected to be emission line stars.
    \item Stars where the template has $T_{\mathrm{eff}}\leq3500\;\mathrm{K}$ or $T_{\mathrm{eff}}\geq7000\;\mathrm{K}$.
\end{itemize}
Additionally, the spectra of all stars where $|RV|>500\;\mathrm{km}\;\mathrm{s}^{-1}$ were visually inspected. A further cut applied in the search by \citet{2018MNRAS.tmp.2466M} for high-velocity stars was to require that \textsc{RV\_NB\_TRANSITS}~$>5$, and this cut has been used widely in the literature. If the measurement is spurious, then it must be spurious in a way that neatly avoids the quality cuts in these earlier works and thus the star must be in some way unusual.

There are two properties of \textit{Gaia} DR2 593...064 that make it slightly unusual among the RVS stars. Firstly, its $G_{\mathrm{RP}}$ magnitude of $13.2$ places it in the faintest $2\%$ of \textit{Gaia} DR2 RVS stars. Secondly, as mentioned in Sec. \ref{sec:scenario1}, \textit{Gaia} DR2 593...064 has the most other \textit{Gaia} sources within $8\;\mathrm{arcsec}$ of the 34 stars that both meet all the criteria above and have $|v_{\mathrm{r}}|>500\;\mathrm{km}\;\mathrm{s}^{-1}$.

The reason to be concerned by the many nearby stars is that it is possible for the light from two stars to blend together in a single RVS spectrum. The \textit{Gaia} RVS is an integral-field spectrograph operating in time delay integration mode, with the result that windows need to be selected around the spectra of individual stars (i.e. there is a conveyor belt of spectra continually being read out). The RVS CCDs see a changing wavelength region for each star as \textit{Gaia} scans across the sky, and thus the overall wavelength scale of each spectra must be determined from the known position of the star from \textit{Gaia} astrometry. When RVS spectra overlap each other on the CCD their windows are truncated in the Across Scan (AC) direction, reducing their AC width down from the nominal 10 pixels.  The truncation is designed to share the flux from the two spectra between the two windows so that each window could be de-blended from the other, however there was no attempt in \textit{Gaia} DR2 to de-blend windows, which means that a single window can contain two different spectra with two different wavelength scales. To mitigate the issue of blended windows, all those windows that were truncated in a non-rectangular pattern were filtered out of the RVS pipeline. Windows that were truncated in a rectangular pattern were let through because most of the time they were truncated by very faint spurious sources. Spurious sources are detected onboard around and along the diffraction spikes of sources brighter than about 16 mag in the SkyMapper CCDs \citep[][not to be confused with the SkyMapper Survey]{2016A&A...595A...3F}.  If the spurious source is brighter than onboard magnitude $G_{\mathrm{RVS}} = 16.2$, then it also gets a RVS window.  RVS windows only start or end on multiples of 105 pixels (before June 2015) or 108 pixels (after June 2015), called macrosample boundaries \citep{2018arXiv180409369C}.  105 and 108 pixels corresponds to approximately 6.2 and 6.4 arcseconds respectively.  Two sources with angular separations in the along scan (AL) direction smaller than these values will have RVS windows starting on the same macrosample boundary.  Approximately 40\% of the stars with onboard magnitude $G_{\mathrm{RVS}}$ between 7 and 9 have a spurious source sufficiently close that the brighter window is truncated in the across scan (AC) direction, such that its AC size is 5 pixels instead of the normal 10, but the two windows are aligned in AL so that the brighter window remains rectangular \citep{2018A&A...616A...6S}. While most truncated windows with rectangular geometries will be due to a spurious source, some spectra containing a contribution from a second bright star will also have been let through.

The implication is that if \textit{Gaia} scans across two stars that are lined up parallel to the AL direction and have an angular offset along the AL direction of less than $6.4\;\mathrm{arcsec}$ (resulting in a rectangular truncation), then their light can be blended into one \textit{Gaia} DR2 RVS spectrum, but with the light from each star offset\footnote{\citet{2018arXiv180409369C} state that the pixels in the along scan direction are $0.0589\;\mathrm{arcsec}$ long and that the dispersion varies from $8.51\;\mathrm{km}\;\mathrm{s}^{-1}\;\mathrm{pix}^{-1}$ at $847\;\mathrm{nm}$ to  $8.58\;\mathrm{km}\;\mathrm{s}^{-1}\;\mathrm{pix}^{-1}$ at $873\;\mathrm{nm}$, which corresponds to $144.5{-}145.7\;\mathrm{km}\;\mathrm{s}^{-1}\;\mathrm{arcsec}^{-1}$ with a mean of $145.1\;\mathrm{km}\;\mathrm{s}^{-1}\;\mathrm{arcsec}^{-1}$.} by $145.1\;\mathrm{km}\;\mathrm{s}^{-1}\;\mathrm{arcsec}^{-1}$, depending on the wavelength under consideration. As mentioned above, this is not a problem in a majority of cases because the other source is usually very faint and spurious. Nevertheless, in a small number of cases the other source may be bright in the $G_{\mathrm{RVS}}$ band and thus could have interfered with the RVS measurement. This interference can range from a small shift in the line centroid to the introduction of a much stronger line which shifts the RVS measurement by hundreds of $\mathrm{km}\;\mathrm{s}^{-1}$. We should expect many of the cases where there are two lines to have been filtered out by the doubly-lined spectroscopic binary cut above.

Contamination of the RVS spectrum of \textit{Gaia} DR2 593...064 by the light of a nearby star can explain the anomalous radial velocity, but would require a star bright in $G_{\mathrm{RVS}}$. The $G_{\mathrm{RVS}}$ was not published in \textit{Gaia} DR2, but, as a proxy, we can say that a star is sufficiently bright to interfere if either it itself has an RVS measurement or is at least as bright as \textit{Gaia} DR2 593...064 in $G_{\mathrm{RP}}$. Indeed, as mentioned previously, \textit{Gaia} DR2 593...352 lies only $4.284\;\mathrm{arcsec}$ away. This corresponds to a velocity offset of $619.0{-}624.2\;\mathrm{km}\;\mathrm{s}^{-1}$. \textit{Gaia} DR2 593...352 has a radial velocity reported in \textit{Gaia} of $5.40\pm2.85\mathrm{km}\;\mathrm{s}^{-1}$, and thus subtracting the velocity offset from the true radial velocity gives $-613.6{-}618.8\;\mathrm{km}\;\mathrm{s}^{-1}$, which encompasses the reported radial velocity for \textit{Gaia} DR2 593...064 of $-614.3\pm2.5\;\mathrm{km}\;\mathrm{s}^{-1}$. The anomalous radial velocity reported for \textit{Gaia} DR2 593...064 can be fully explained if the spectra used to determine the radial velocity was blended with that of \textit{Gaia} DR2 593...352 in each of the seven radial velocity transits.

\begin{figure}
	\includegraphics[width=\linewidth,clip, trim=2.0cm 0.0cm 2.8cm 1.2cm]{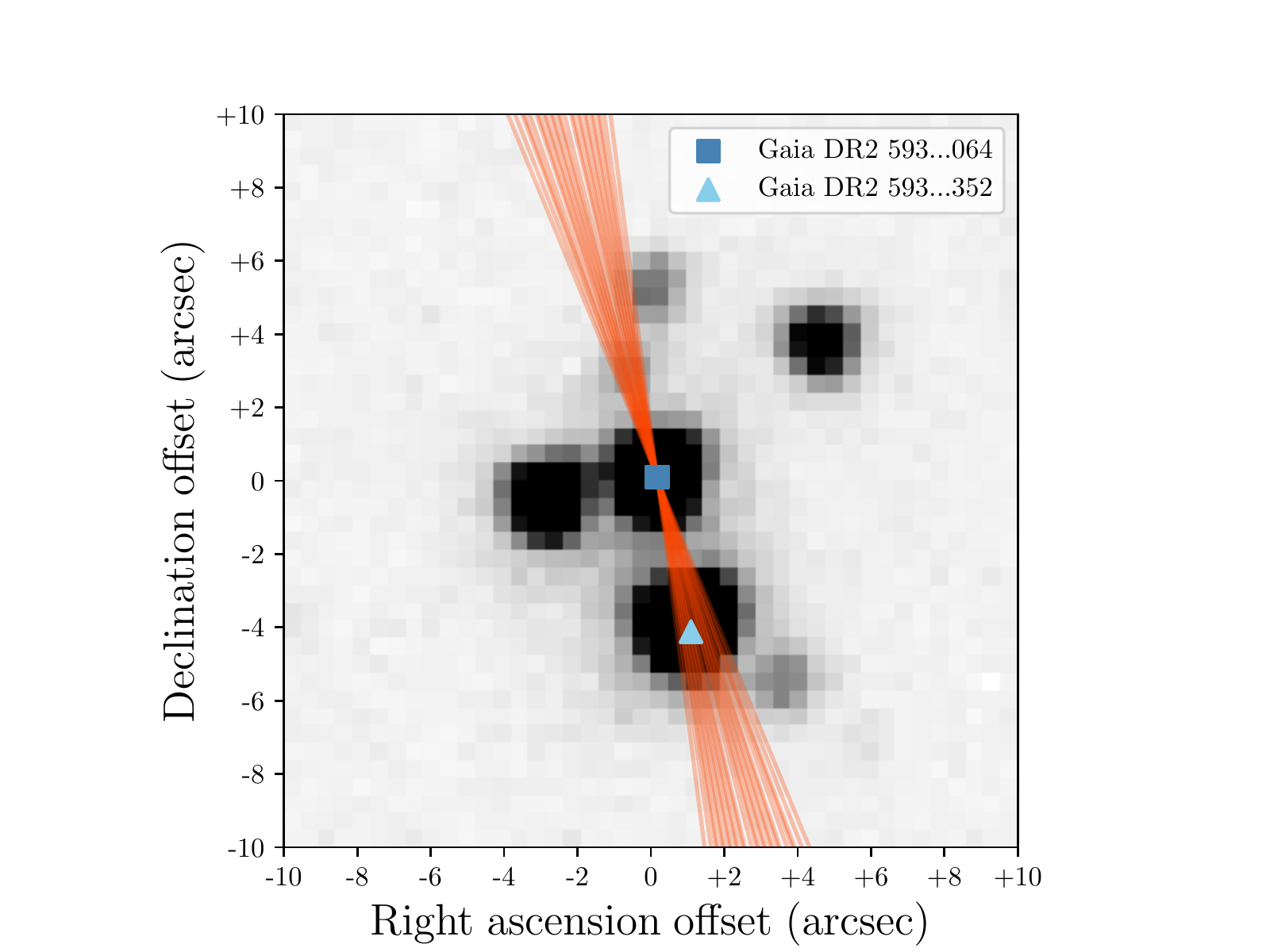}
    \caption{The seven radial velocity measurements of \textit{Gaia} DR2 593...064 were likely made during a subset of the 26 scans that are shown as orange lines (fixed to pass through the centre of that star). These scans all pass through the nearby bright star \textit{Gaia} DR2 593...352 and thus it is likely that all seven RVS spectra are blends of the light from both stars. Note that the SkyMapper image has been flipped with respect to Fig. \ref{fig:cutouts}.}
    \label{fig:gaiatracks}
\end{figure}

The scenario outlined above requires that most of the seven radial velocity transits occurred during \textit{Gaia} scans that passed across both stars. The Gaia Observation Forecasting Tool (GOST) provides both the dates and scanning angles of the possible radial velocity measurements, however, as discussed above, we do not know which 7 of these 42 possibilities actually contributed to the RVS measurement. The small standard deviation of the radial velocity measurements ($5.17\;\mathrm{km}\;\mathrm{s}^{-1}$) provides one clue: if even a single transit gave a measurement of the true radial velocity of \textit{Gaia} DR2 593...064 then the standard deviation would be much greater. Thus, looking at Fig. \ref{fig:rvdates}, it seems quite likely that all seven transits will have occurred in the 3.75-day window. In Fig. \ref{fig:gaiatracks}, we show the 26 \textit{Gaia} scans that occurred over that time period. These scans all pass through the nearby bright star \textit{Gaia} DR2 593...352 and thus it is likely that all seven RVS spectra are blends of the light from both stars.

We conclude that it is highly likely that the radial velocity measurement reported in \textit{Gaia} DR2 is spurious and thus that \textit{Gaia} DR2 593...064 has a systemic velocity along the line-of-sight of $-62.5\pm5.0\;\mathrm{km}\;\mathrm{s}^{-1}$. Assuming the system has a radial velocity of $-62.5\;\mathrm{km}\;\mathrm{s}^{-1}$ with proper motion $(\mu_{\alpha}\cos{\delta},\mu_{\delta})=(-2.676,-4.991)\;\mathrm{mas}\;\mathrm{yr}^{-1}$ at a distance of $2.08\;\mathrm{kpc}$, we find that \textit{Gaia} DR2 593...064 is a typical disk star with position $(R,z)=(6.3,-0.1)\;\mathrm{kpc}$ and velocity $(v_R,v_{\phi},v_z)=(33,288,-7)\;\mathrm{km}\;\mathrm{s}^{-1}$. It is likely that \textit{Gaia} DR2 593...064 is co-moving with \textit{Gaia} DR2 593...576, further strengthening this conclusion.

\subsection{Implications for \textit{Gaia} DR2 radial velocities}
\label{sec:implications}

If the cuts described in the previous section failed to catch the spurious measurement of \textit{Gaia} DR2 593...064, then it is possible that there are more stars that suffer similar issues. For each star in \textit{Gaia} RVS, we query in the full \textit{Gaia} DR2 catalogue whether there is another source within $6.4\;\mathrm{arcsec}$ that either has an RVS measurement or is at least as bright in $G_{\mathrm{RP}}$. This applies to 63764 of the \textit{Gaia} RVS stars. Many of these pairings will not have resulted in blended spectra, but because of the limited data published in \textit{Gaia} DR2 and the infeasibility of subjecting every one of these sources to the same treatment we have performed on \textit{Gaia} DR2 593...064, we recommend cutting these sources.


One complication is that the \textit{Gaia} $G_{\mathrm{BP}}$ and $G_{\mathrm{RP}}$ pipeline was more conservative in their treatment of blends than the RVS pipeline, and discarded the colours of stars where blends were suspected. A star not having a $G_{\mathrm{RP}}$ measurement can thus be suggestive of a blend. Note that \textit{Gaia} DR2 593...352 is one of the  10976 \textit{Gaia} RVS sources that are missing one or both of $G_{\mathrm{BP}}$ and $G_{\mathrm{RP}}$. We recommend cutting all RVS sources missing either of these photometric bands. We also extend the neighbour blending criteria above to encompass nearby stars that are at least as bright in $G$, which increases the number of suspect \textit{Gaia} RVS stars to 70365 stars.

Another way to flag \textit{Gaia} DR2 radial velocities as not representative of the systemic velocity of the system or suspicious is if the variance across the individual radial velocity measurements is unusual. We expect that the variance should be primarily a function of $G_{\mathrm{RP}}$ and $G_{\mathrm{BP}}-G_{\mathrm{RP}}$ (for instance, hot stars will have smaller and broader lines), and thus an usually large or small variance could indicate that the radial velocity cannot be trusted. An unusually large variance could be caused by the star being in a binary; singly-lined spectroscopic binaries were not treated in the \textit{Gaia} DR2 RVS pipeline. Alternatively, if a star has an unusually small variance then that could signal a star whose spectrum was dominated by the light of a nearby brighter star, because then the uncertainty on the radial velocity will be the relatively smaller uncertainty of a brighter star. We defer the proper investigation of the \textit{Gaia} DR2 radial velocity variances to a forthcoming paper (Boubert et al., in prep.), however we do suggest that stars with too few RV transits should be removed based on the following argument.

Suppose that \textit{Gaia} makes $N$ radial velocity measurements of a star and that those radial velocity measurements are drawn from a Gaussian centred on the true mean radial velocity with a standard deviation of $\sigma$. It then follows that the variance $S^2=S(v_r^t)^2$ of those radial velocity measurements will have as its sampling distribution
\begin{equation}
\frac{(N-1)S^2}{\sigma^2} \sim \chi^2_{N-1}. \label{eq:chi2}
\end{equation}
This equation implies that the likelihood of observing a radial velocity variance $S^2$ given the number of observations $N$ and true variance $\sigma^2$,
\begin{equation}
\operatorname{Likelihood}\left(S^2|N,\sigma^2\right) = \frac{(N-1)}{\sigma^2} \chi^2_{N-1}\left(\frac{(N-1)S^2}{\sigma^2}\right). \label{eq:chi2}
\end{equation}
The value of $\sigma^2$ that maximizes this likelihood is $S^2$ for all $N>1$. Note, however, that this likelihood is not a well-defined probability density function for $N<4$, because its integral does not converge over the region $(0,\infty)$. While we can resolve this by specifying a prior on $\sigma^2$ and using the Bayesian posterior rather than maximum likelihood estimation to obtain a most likely value, this improper behaviour suggests that any constraint we infer on $\sigma^2$ for stars with $N<4$ will be too broad to rule on whether such stars are single or binary. We therefore recommend an additional quality cut that requires the number of transits $N\geq4$. While the justification above is qualitative in nature, there is a further reason to cut stars with few transits: a cut on $N$ acts to remove stars with blended spectra, because the probability that the $N$ transits are sufficiently aligned decreases with increasing $N$.

In summary, we recommend three novel quality cuts that could be applied to the \textit{Gaia} RVS sample. We list these here in order of increasing complexity:
\begin{itemize}
	\item The star must have reported $G_{\mathrm{BP}}$ and $G_{\mathrm{RP}}$ magnitudes.
	\item The radial velocity must be based on at least four transits (e.g. \textsc{rv\_nb\_transits}$\;\geq4$).
	\item The star must not have a neighbour within $6.4\;\mathrm{arcsec}$ that either itself is in RVS or that is brighter in $G$ or $G_{\mathrm{RP}}$.
\end{itemize}
The first and the third cuts aim to exclude sources with radial velocities that are suspected to be contaminated.  This paper finds one source that is highly likely to be contaminated but it is possible that these cuts will also exclude uncontaminated sources with valid radial velocities.  The first and the third cuts will obtain the cleanest possible sample (in terms of contamination) but at the expense of completeness.  The second cut removes sources where the number of transits are too few to determine whether the radial velocity is representative of the systemic velocity of the system.  This may remove sources where the radial velocities are actually representative of the systemic velocity of the system.  Similarly to the other cuts, the second cut will obtain the cleanest possible sample (in terms of well-behaved radial velocity variances) but at the expense of completeness.  Therefore, the applicability of these cuts depends on the science question being addressed.

\begin{figure*}
	\includegraphics[width=\linewidth,clip, trim=1.7cm 0.5cm 2.7cm 1.5cm]{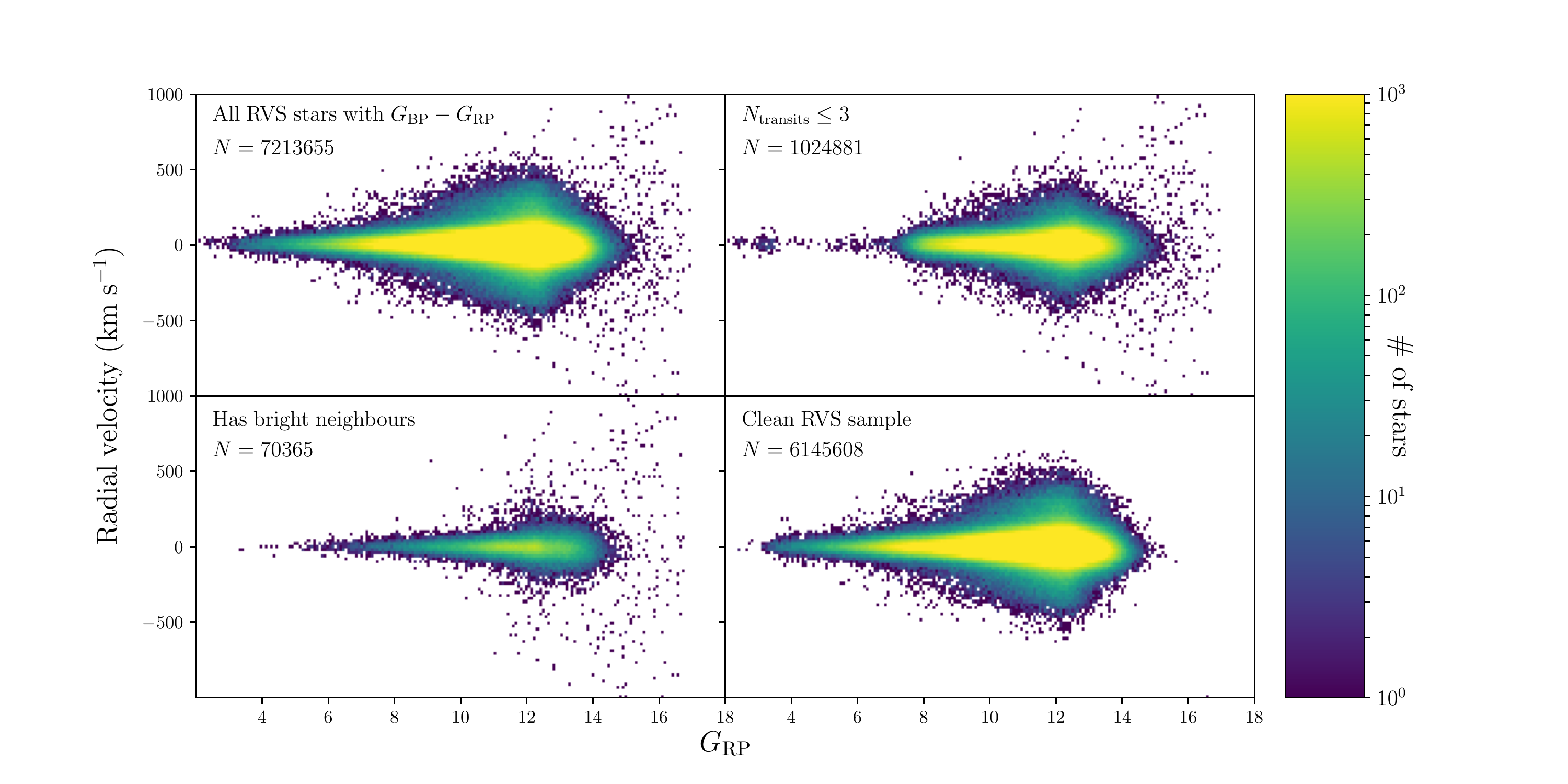}
	\caption{Each panel shows the 2D histogram of a different selection of \textit{Gaia} RVS stars. The top left panel shows all the stars, while the bottom right shows the distribution after applying the quality cuts described in Sec. \ref{sec:implications}.}
	\label{fig:rvscontamination}
\end{figure*}

We demonstrate the efficacy of these cuts in Fig. \ref{fig:rvscontamination}. The final clean sample contains 6145608 stars, which is many more than the 4809107 that survive the commonly used \textsc{rv\_nb\_transits}$\;>5$ cut. Of the 202 stars with radial velocities greater than $500\;\mathrm{km}\;\mathrm{s}^{-1}$, only 90 survive the four cuts. To aid the reader in implementing the nearby, bright companion cut, we give the necessary information in Tab. \ref{tab:catalogue}.

\begin{table}
 \caption{List of all sources in \textit{Gaia} RVS which have a companion in the full \textit{Gaia} DR2 catalogue within $6.4\;\mathrm{arcsec}$ that either itself is in \textit{Gaia} RVS or that is brighter in $G_{\mathrm{RP}}$ or $G$. The columns are the \textit{Gaia} DR2 ID, the number $N_{\mathrm{RVS}}$ of companions that are in RVS, the number $N_{\mathrm{RP}}$ of companions brighter in $G_{\mathrm{RP}}$ and the number $N_{\mathrm{G}}$ of companions brighter in $G$. Only the first five rows are shown here and the full table is available in a supplementary datafile online.}
 \label{tab:catalogue}
 \centering
 \begin{tabular}{rrrr}
  \hline
  SourceID & $N_{\mathrm{RVS}}$ & $_{\mathrm{\phantom{S}}}N_{\mathrm{RP}}$ & $_{\mathrm{\phantom{VS}}}N_{\mathrm{G}}$\\
  \hline
   83154861954304 & 0 & 1 & 1 \\
  739666383070976 & 0 & 1 & 0 \\
  969837975192832 & 1 & 0 & 0 \\
  969842270721536 & 1 & 1 & 1 \\
  990629912562048 & 1 & 0 & 0 \\
  \hline
 \end{tabular}
\end{table}

This work investigated the most likely of the hypervelocity candidates proposed by \citet{2018MNRAS.tmp.2466M}, \citet{2018ApJ...866..121H} and \citet{2018ApJ...868...25B}, however we checked whether the issue of unreliable \textit{Gaia} DR2 radial velocities affects any of the other candidates in the literature. We cross-matched The Open Fast Stars Catalog \citep{2018MNRAS.479.2789B} against our list of stars with unreliable radial velocities and identified two additional candidates which may suffer from these issues. 
\textit{Gaia} DR2 5958197813784543872 (otherwise known as 2MASS 17464606-3937523) was proposed by \citet{2012AJ....143...57K} as a possible hypervelocity star based on their measured $447\;\mathrm{km}\;\mathrm{s}^{-1}$, although those authors commented that its metallicity of $[\mathrm{Fe}/\mathrm{H}]=-0.86\pm0.05$ is consistent with it being a bound bulge star. This star was flagged because the median \textit{Gaia} DR2 radial velocity of $421.63\pm3.09\;\mathrm{km}\;\mathrm{s}^{-1}$ is based on only two measurements, however we note that this radial velocity is consistent with the value reported by \citet{2012AJ....143...57K}. The other flagged star is \textit{Gaia} DR2 5300505902646873088 (otherwise known as Gaia-T-ES2) which is a $G_{\mathrm{RP}}=12.4$ star ranked by \citet{2018ApJ...866..121H} as the second most likely unbound candidate in their list. The \textit{Gaia} radial velocity of $160.22\pm4.00\;\mathrm{km}\;\mathrm{s}^{-1}$ was flagged because it is based on only three measurements.

\section{Conclusions}

\textit{Gaia} DR2 593...064 was proposed by \citet{2018MNRAS.tmp.2466M} and \citet{2018ApJ...868...25B} to be a likely hypervelocity star based on the incredible radial velocity $-614.3\pm2.5\;\mathrm{km}\;\mathrm{s}^{-1}$ reported in \textit{Gaia} DR2. \citet{2018ApJ...868...25B} suggested that this star may be in a crowded field due to its proximity to the Galactic plane; taking images from SkyMapper, we found that the star is indeed in a crowded field surrounded by several relatively bright stars. Motivated by this, we obtained eight epochs of ground-based spectroscopic follow-up with the SOAR Telescope. From these spectra, we measured a median radial velocity of $-56.5 \pm 5.3\;\mathrm{km}\;\mathrm{s}^{-1}$, which is seemingly inconsistent with the radial velocity reported in \textit{Gaia} DR2. Analysis of the eight spectra determined that \textit{Gaia} DR2 593...064 is an A-type main-sequence or sub-giant star, and that the star is not spectroscopically unusual in a way that could explain the discrepancy.

The \textit{Gaia} measurement is based on seven individual radial velocity measurements taken from seven RVS transits, each transit corresponding to three RVS CCD spectra. Neither the individual measurements nor the dates on which they were taken are publicly available, however it seems likely that some of the seven measurements were taken during a 3.75-day window beginning on $7^{\mathrm{th}}$ July 2015. This allowed us to infer that if the \textit{Gaia} measurement is correct, then the star must be in orbit around an intermediate-mass black hole, which suggests that the \textit{Gaia} datum is likely spurious. By contrast, the eight radial velocities we obtained show evidence for binary motion with a period of less than $70\;\mathrm{days}$ and we constrained the likely parameters of that binary.

The spurious \textit{Gaia} radial velocity is most probably caused by light from a nearby bright star blending with the spectrum of \textit{Gaia} DR2 593...064. The \textit{Gaia} RVS is an integral field spectrograph that operates in time delay integration mode, and thus the angular offset between the two stars translates into a velocity shift of the contaminating spectrum relative to that of \textit{Gaia} DR2 593...064 at a rate of $145.1\;\mathrm{km}\;\mathrm{s}^{-1}\;\mathrm{arcsec}^{-1}$. This effect relies on all seven scans passing across both stars, and thus makes it highly likely that all seven transits occurred in the 3.75-day window beginning on $7^{\mathrm{th}}$ July 2015.

That the reported radial velocity of \textit{Gaia} DR2 593...064 could be so badly wrong begs the question: how many other \textit{Gaia} DR2 RVS sources are susceptible? We find that any star with a companion closer than $6.4\;\mathrm{arcsec}$ that either itself has an RVS measurement or is brighter in $G$ or $G_{\mathrm{RP}}$ could be suspect. For the cleanest possible sample (in terms of contamination), we also recommend that RVS stars without $G_{\mathrm{BP}}$ or $G_{\mathrm{RP}}$ should be cut, because one reason for the absence of colour photometry is if the pipeline detected a blend in the BP/RP spectra. The radial velocity variances of single \textit{Gaia} RVS stars should be a simple function of $G_{\mathrm{RP}}$ and $G_{\mathrm{BP}}-G_{\mathrm{RP}}$, and thus stars with an excessively large radial velocity variance are likely in a singly-lined spectroscopic binary (a possibility that is not treated in the RVS pipeline for \textit{Gaia} DR2). While we leave a full analysis of the radial velocity variances to a forthcoming work (Boubert et al., in prep.), we argue that the radial velocity variance of stars with three or fewer transits cannot provide strong enough evidence that the radial velocity variance is well-behaved, and thus, for the cleanest possible sample (in terms of well-behaved radial velocity variances), we recommend that only stars with four or more transits are used. These cuts are effective: they remove almost all of the sources which have large radial velocities or are too faint to have had a true RVS measurement in DR2, whilst retaining 85\% of the stars.

Many of the issues that our three cuts target will be handled by the RVS pipeline in future \textit{Gaia} data releases. However, at the time of writing, the third \textit{Gaia} data release will not be until the first half of 2021\footnote{\url{https://www.cosmos.esa.int/web/gaia/release} accessed on 23/11/2018.}. The cuts presented in this work will unlock the full potential of \textit{Gaia} DR2 radial velocities and open the door to progress in Galactic dynamics over the two years prior to DR3.

\section*{Acknowledgements}

The authors would like to thank Morgan Fraser, David Katz, Payel Das, Keith Hawkins and Lachlan Lancaster for useful discussions. This work is an offshoot of the D$^6$ collaboration of theorists and observers hunting for hypervelocity white dwarfs. DB thanks Magdalen College for his Fellowship by Examination. DA is supported by the Leverhulme Trust, whilst JLS holds a Leverhulme/Isaac Newton Trust Fellowship. JS acknowledges support from the Packard Foundation. SK is partially supported by NSF grant AST-1813881. Based on observations obtained at the Southern Astrophysical Research (SOAR) telescope, which is a joint project of the Minist\'{e}rio da Ci\^{e}ncia, Tecnologia, Inova\c{c}\~{o}es e Comunica\c{c}\~{o}es (MCTIC) do Brasil, the U.S. National Optical Astronomy Observatory (NOAO), the University of North Carolina at Chapel Hill (UNC), and Michigan State University (MSU). This work has made use of data from the European Space Agency (ESA) mission {\it Gaia}(\url{https://www.cosmos.esa.int/gaia}), processed by the {\it Gaia} Data Processing and Analysis Consortium (DPAC, \url{https://www.cosmos.esa.int/web/gaia/dpac/consortium}). Funding for the DPAC has been provided by national institutions, in particular the institutions participating in the {\it Gaia} Multilateral Agreement. This research has made use of ESASky, developed by the ESAC Science Data Centre (ESDC) team and maintained alongside other ESA science mission's archives at ESA's European Space Astronomy Centre (ESAC, Madrid, Spain). This research made use of Astropy,\footnote{http://www.astropy.org} a community-developed core Python package for Astronomy \citep{astropy:2013, astropy:2018}, as well as the \textsc{Python} packages \textsc{corner} \citep{corner}, \textsc{emcee} \citep{emcee} and \textsc{The Joker} \citep{Price-Whelan:2017a,2017ApJ...837...20P}.




\InputIfFileExists{main.bbl}



\appendix

\section{Posterior corner plot for atmospheric and stellar parameters}
\label{sec:atmoscorner}

\begin{figure*}
	\includegraphics[width=\linewidth]{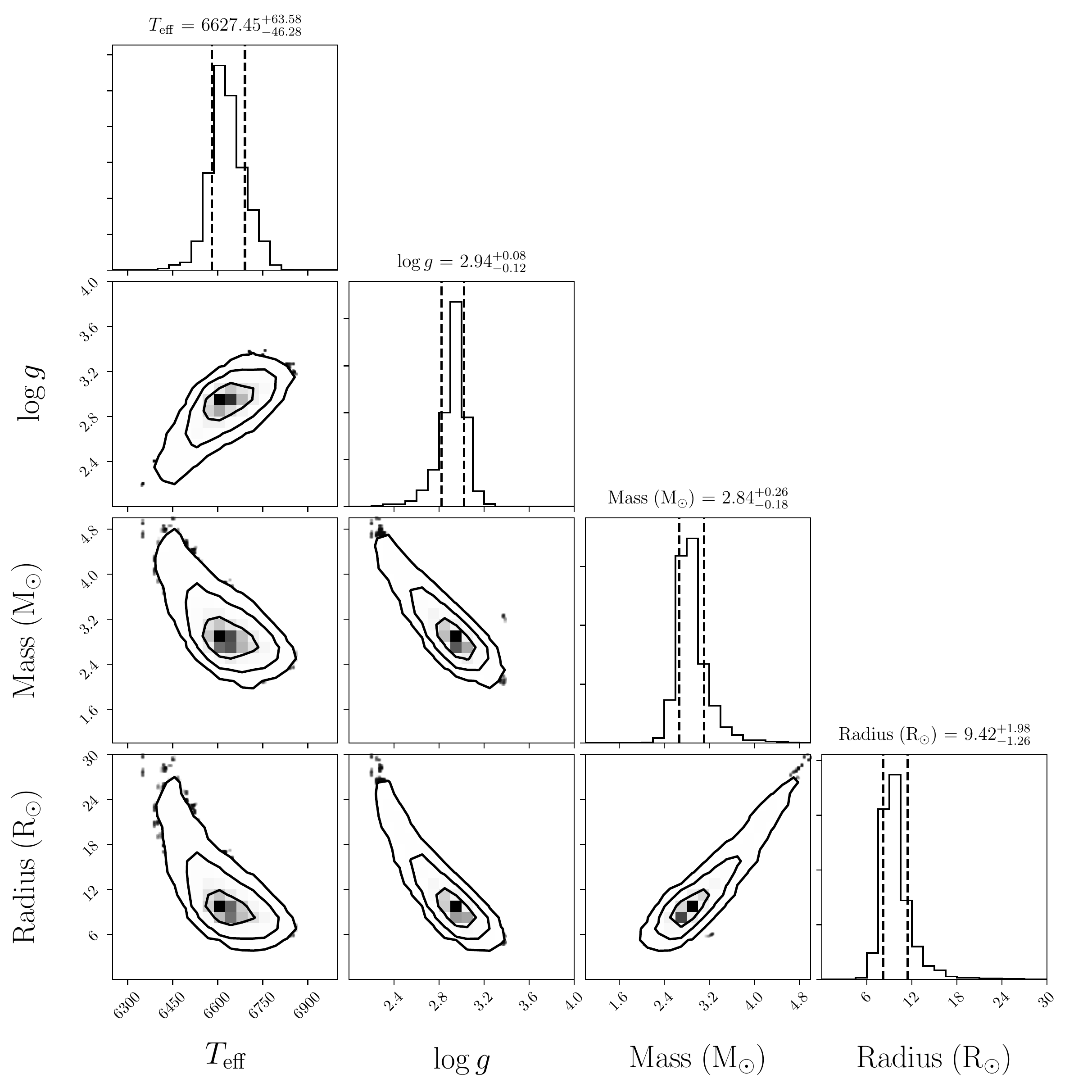}
    \caption{Corner plot of the posterior samples of the effective temperature $T_{\mathrm{eff}}$, surface gravity $\log{g}$, mass and radius of \textit{Gaia} DR2 593...064. The procedure that resulted in this plot is described in detail in Sec. \ref{sec:observations}.}
    \label{fig:speciso}
\end{figure*}


\bsp	
\label{lastpage}
\end{document}